\documentstyle[prl,aps,psfig,multicol]{revtex}
\begin{document}
\draft
\title{Reflection symmetry at a $B=0$ metal-insulator transition in two dimensions}
\author{D.\ Simonian, S.\ V.\ Kravchenko, and M.\ P.\ Sarachik}
\address{Physics Department, City College of the City University of New York, 
New York, New York 10031}
\date{\today}
\maketitle
\begin{abstract}
We report a remarkable symmetry between the resistivity and conductivity on
opposite sides of the $B=0$ metal-insulator transition in a two-dimensional
electron gas in high-mobility silicon MOSFET's. This symmetry implies that the transport mechanisms on the two sides are related.
\end{abstract}
\begin{multicols}{2}

Within the scaling theory of localization\cite{gang} developed for 
non-interacting electrons, no metallic phase exists in two dimensions in the 
absence of a magnetic field and no metal-insulator transition is therefore 
possible.  Contrary to this expectation, several recent 
experiments\cite{krav,surface,efield} have given clear indication of a 
metal-insulator transition in zero magnetic field 
in a two-dimensional electron gas in high-mobility silicon 
metal-oxide-semiconductor field-effect transistors (MOSFET's).  Measurements 
in 
samples equipped both with aluminum\cite{krav,efield} and polysilicon 
\cite{surface} gates have demonstrated that the 2D gas of electrons exhibits 
behavior that is characteristic of a true phase transition: the resistivity 
scales with temperature\cite{krav,surface} and electric field\cite{efield} with 
a single parameter that approaches zero at a critical electron density $n_c$.  
The nature of 
this unexpected transition and the physical mechanism that drives it are not 
understood.

In GaAs/AlGaAs heterostructures, Shahar {\it et al.} \cite{shahar1} have 
recently found a direct and simple relation between the longitudinal 
resistivity in the magnetic field-induced insulating phase and the 
neighboring quantum Hall liquid (QHL) phase: $\rho_{xx} (\Delta \nu) = 
1/\rho_{xx} (-\Delta \nu)$.  Here $\Delta \nu = \nu - \nu_c$, and $\nu_c$ is 
the critical filling factor for the $\nu = 1$ QHL-insulator transition; the 
relation also holds for the fractional $\nu = 1/3$ QHL-insulator transition 
when mapped \cite{jain} onto the $\nu = 1$ QHL-insulator transition of 
composite Fermions.  Shahar {\it et al.} \cite{shahar1} point out that this 
remarkable symmetry indicates a close relation between the conduction 
mechanisms in the two phases.

In this paper, we report a similar symmetry near the critical electron 
density for the $B=0$ metal-insulator transition in the 2D electron gas in 
high mobility silicon MOSFET's.  Over a range of 
temperature $0.3\text{ K}<T<1\text{ K}$, the (normalized) linear conductivity 
on either side of the transition is equal to its inverse on the other side:
\begin{equation}
\rho^*(\delta_n,T)=\sigma^*(-\delta_n,T).\label{symm}
\end{equation}
Here $\delta_n\equiv (n_s - n_c)/n_c$, $n_s$ is the electron density, $n_c$ is 
the 
critical electron density, $\rho^*\equiv\rho/\rho_c$ is the resistivity 
normalized by its value, $\rho_c\approx3h/e^2$, at the transition, and 
$\sigma^*\equiv1/\rho^*$.
In the case of the magnetic field-induced QHL-insulator transition, the 
symmetry was attributed to charge-flux duality \cite{shahar}.  The 
observation of similar behavior in a 2D electron gas in the absence of a 
magnetic field implies that flux does not play a role in this case.  Although 
the observed duality may have different 
\vbox{
\hbox{
\hspace{0.5in}
\psfig{file=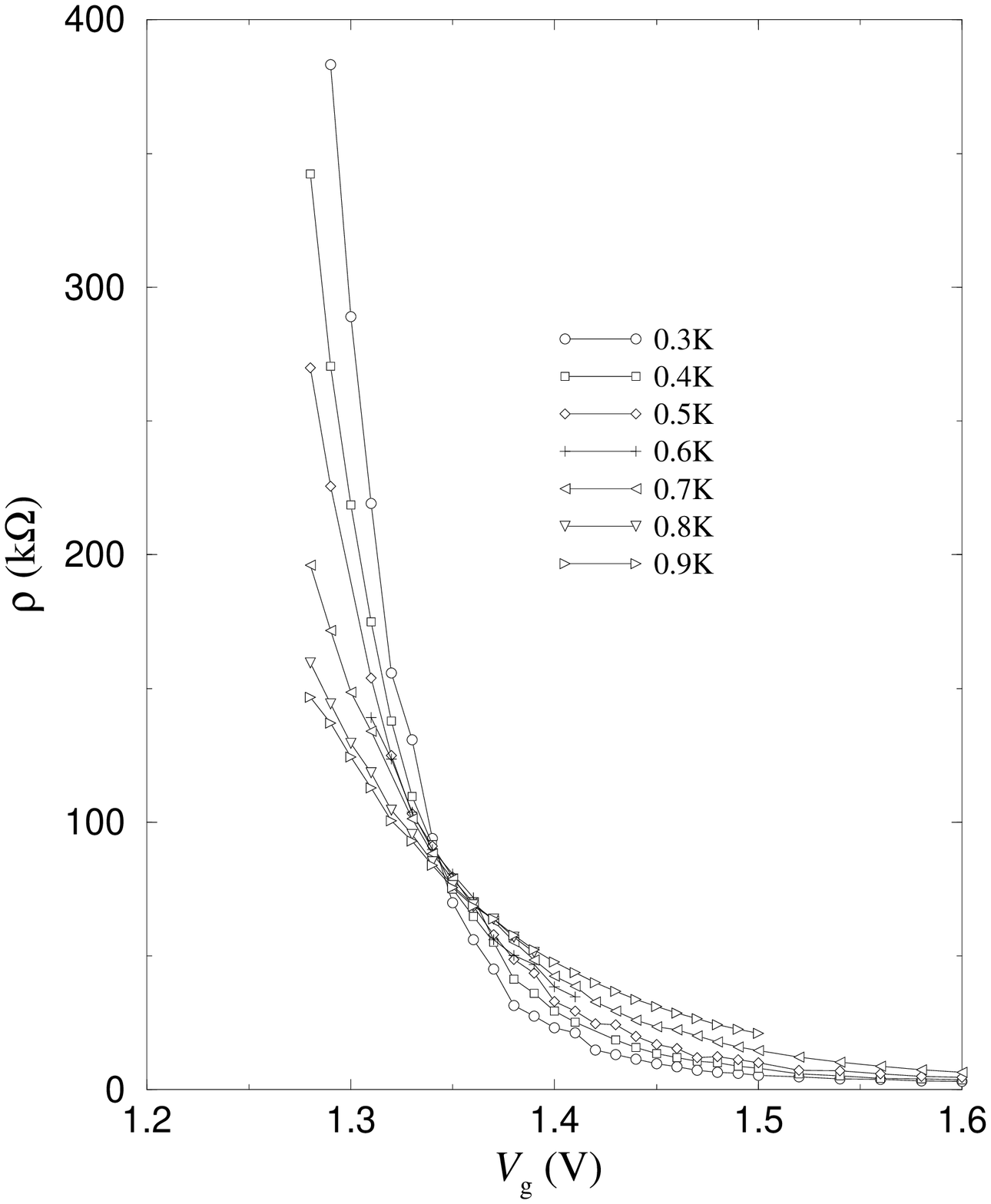,width=3.0in,bbllx=1.5in,bblly=1in,bburx=7.75in,bbury=9.25in,angle=0}
}
\vspace{0.15in}
\hbox{
\hspace{-0.15in}
\refstepcounter{figure}
\parbox[b]{3.4in}{\baselineskip=12pt \egtrm FIG.~\thefigure.
Resistivity as a function of gate voltage, $V_g$, for temperatures between 
$0.3\text{ K and } 0.9\text{ K}$, obtained from the linear portion of the 
$I-V$ curves using the appropriate dimensionless geometric factor.\vspace{2mm}
}
\label{fig1}
}
}
underlying causes, our results suggest that it may originate with some 
fundamental feature that is common to 
both.
 
Four terminal DC resistivity measurements were performed on high quality 
silicon MOSFET's with maximum electron mobilities 
$\mu^{max}\approx35,000-40,000\text{ cm}^2/\text{Vs}$ similar to the samples 
used in Refs.\cite{krav,efield}.  Different electron densities were obtained 
in the usual manner by controlling the gate voltage, $V_g$.  $I-V$  curves 
were recorded at each temperature and electron density, and the resistivity 
was determined from the slope of the linear portion of the curve.

Fig.~\ref{fig1} shows the resistivity as a function of gate voltage (electron 
density) at several different temperatures between $0.3\text{ K}$ and 
$0.9\text{ K}$.  The curves all intersect at a single value of the gate 
voltage, $V_g=1.348$~V, corresponding to a critical electron density, 
$n_c=8.45\times10^{10}$~cm$^{-2}$.   The 
\vbox{
\hbox{
\hspace{0.10in}
\psfig{file=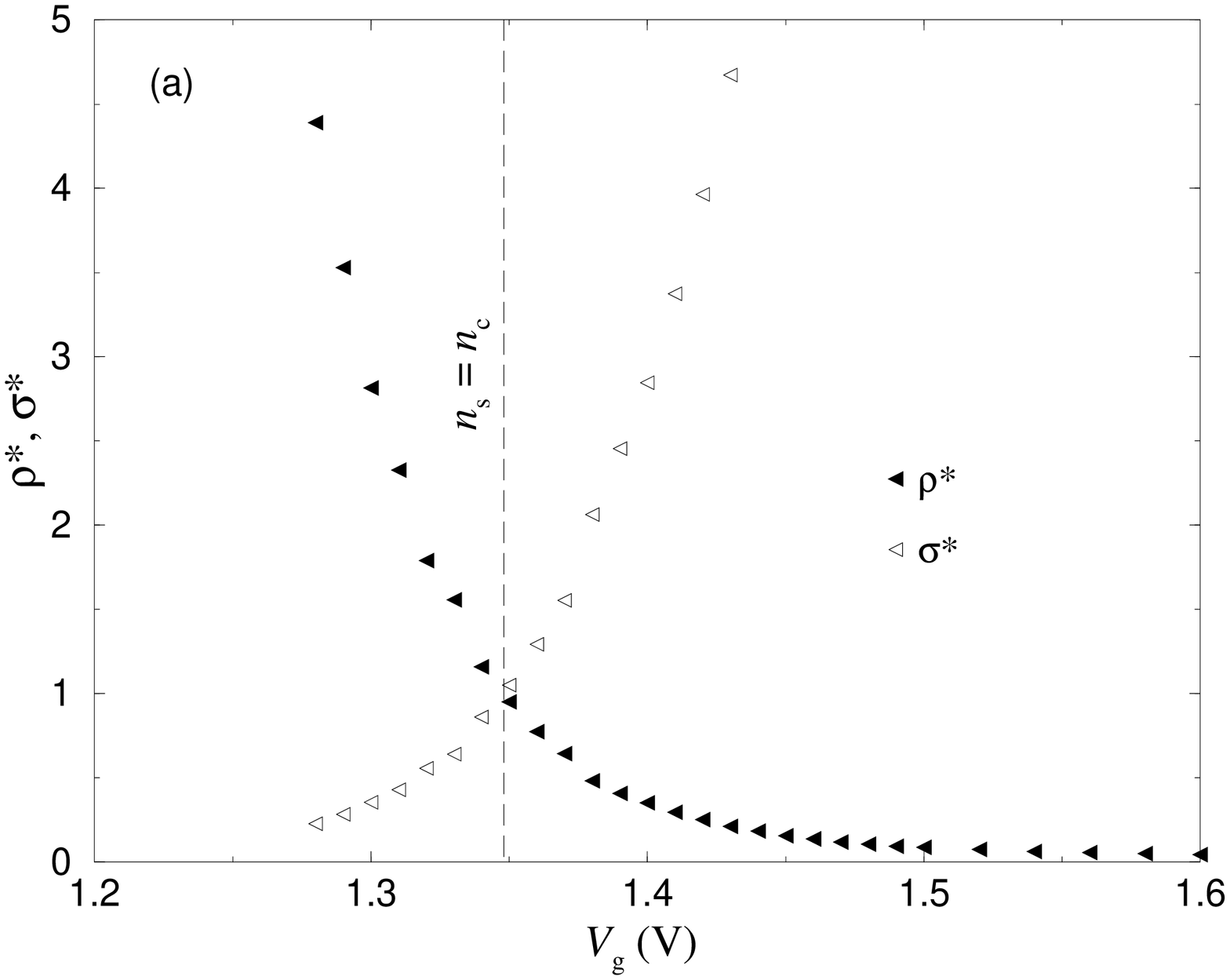,width=3.0in,bbllx=.5in,bblly=1.25in,bburx=7.25in,bbury=9.5in,angle=-90}
}
}
\vbox{
\hbox{
\hspace{0.10in}
\psfig{file=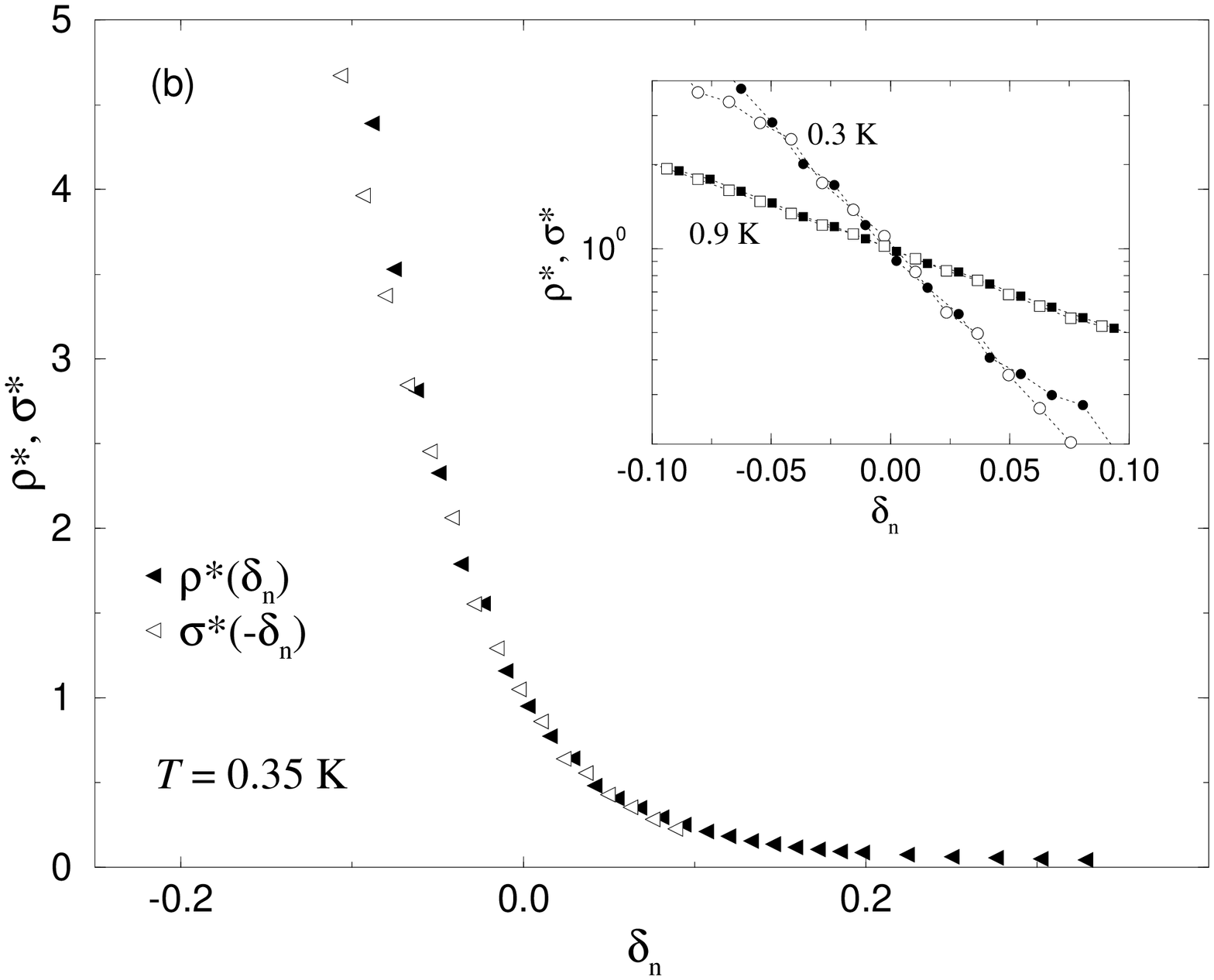,width=3.0in,bbllx=.5in,bblly=1.25in,bburx=7.25in,bbury=9.5in,angle=-90}
}
\vspace{0.3in}
\hbox{
\hspace{-0.15in}
\refstepcounter{figure}
\parbox[b]{3.4in}{\baselineskip=12pt \egtrm FIG.~\thefigure.
(a)~Normalized resistivity, $\rho^*$, and normalized conductivity, $\sigma^*$, 
as functions of the gate voltage, $V_g$, at $T=0.35$~K. Note the symmetry about 
the line $n_s = n_c$. The electron density is given by $n_s= 
(V_g-0.58\text{V})\times1.1\times10^{11}$~cm$^{-2}$.
 (b)~To demonstrate this symmetry explicitly, $\rho^*(\delta_n)$ (closed 
symbols) and $\sigma^*(-\delta_n)$ (open symbols) are plotted versus 
$\delta_n\equiv(n_s-n_c)/n_c$. Inset: $\rho^*(\delta_n)$ (closed symbols) and  
$\sigma^*(-\delta_n)$ (open symbols) versus $\delta_n$ at $T=0.3$~K and 
$T=0.9$~K, the lowest and highest measured temperatures.
\vspace{0.10in}
}
\label{fig2}
}
}
resistivity decreases 
(increases) with increasing temperature for $n_s<n_c$ ($n_s>n_c$), as 
expected for insulating (metallic) behavior.  In agreement with earlier 
measurements\cite{krav,surface,efield}, the resistivity at the critical point 
is close to $3h/e^{2}$.

The normalized resistivity $\rho^*(V_g)$ and the normalized conductivity 
$\sigma^*(V_g)$ at $T=0.35$~K are shown as functions of the gate voltage in 
Fig.~\ref{fig2}~(a).  Note the apparent symmetry about the vertical line 
corresponding to the critical electron density.  Fig.~\ref{fig2}~(b) 
demonstrates that the curves can be mapped onto each other by reflection, {\it 
i.e.}, $\rho^*(\delta_n)$ is virtually identical to $\sigma^*(-\delta_n)$.  
Our data indicate that this mapping holds over a range 
of temperature from $0.3\text{ K}$ to $0.9\text{ K}$. However, the range 
$|\delta_n|$ over which it holds decreases continuously as the temperature 
\vbox{
\vspace{0.35in}
\hbox{
\hspace{0.1in}
\psfig{file=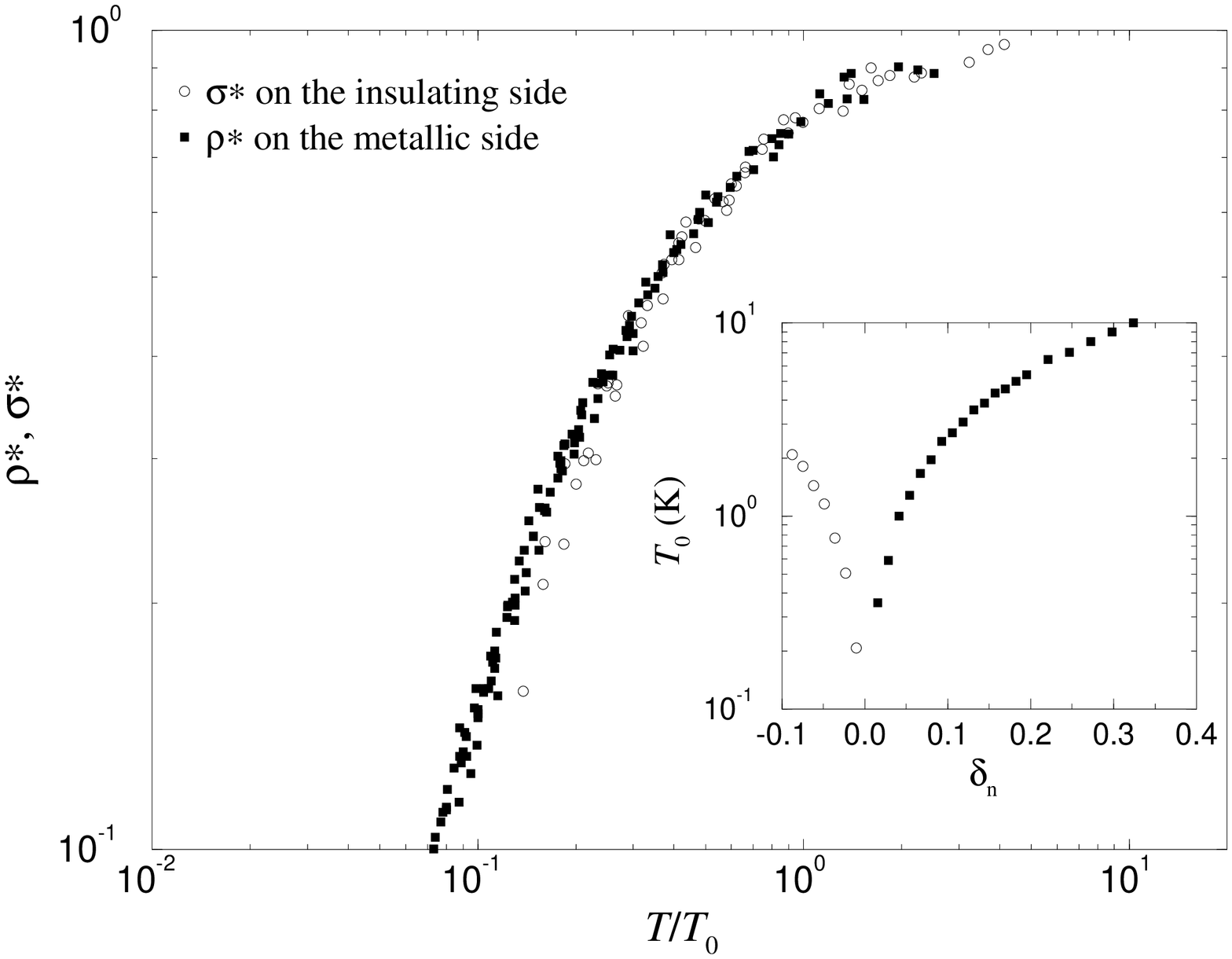,width=3.0in,bbllx=1.5in,bblly=1in,bburx=7.75in,bbury=9.25in,angle=-90}
}
\vspace{0.15in}
\hbox{
\hspace{-0.15in}
\refstepcounter{figure}
\parbox[b]{3.4in}{\baselineskip=12pt \egtrm FIG.~\thefigure.
Normalized resistivity $\rho^*$ on the metallic side of the transition (closed 
symbols) and normalized conductivity $\sigma^*$ on the insulating side (open 
symbols) versus scaled temperature, $T/T_0$. The 
scaling parameter $T_0$ is shown as a function of $\delta_n$ in the inset.
\vspace{0.10in}
}
\label{fig3}
}
}
is decreased:  
for example, at $T=0.9\text{ K}$, $\rho^*$ and $\sigma^*$ are symmetric for 
$|\delta_n|\lesssim 0.1$, while at $T=0.3\text{ K}$, they are symmetric only 
for $|\delta_n|\lesssim 0.05$ (see inset to Fig.~\ref{fig2}~(b)).

The resistivity of the 2D electron gas in Si MOSFET's was shown\cite{krav} to 
scale near the transition according to:
\begin{equation}
\rho(T,\delta_n) = f(|\delta_n|/T^b) = \rho(T/T_0)
\label{tscal}
\end{equation}
with a single parameter $T_0$ that is the same function of $|\delta_n|$ on 
both the metallic and the insulating side of the transition, $T_0\propto 
|\delta_n|^{1/b}$.  Combined with the scaling of Eq.(\ref{tscal}), the 
duality expressed in Eq.(\ref{symm}) takes the form
\begin{equation}
\rho_{\text{met,ins}}^*(T/T_0) = \sigma_{\text{ins,met}}^*(T/T_0).
\label{scalsymm}
\end{equation}
The scaled curves $\rho^*(T/T_0)$ for the metallic side and $\sigma^*(T/T_0)$ 
for the insulating side should thus be equivalent.  This is demonstrated in 
Fig.~\ref{fig3}, which shows that $\rho^*(T/T_0)$ and $\sigma^*(T/T_0)$ are 
indeed virtually identical for a given sample in a range where the 
resistivity (conductivity) changes by an 
order of magnitude. Remarkably, this indicates that the temperature dependence 
of the resistivity in 
either phase is similar to the temperature dependence of the conductivity 
in the other phase, implying that the mechanisms responsible for electrical 
transport in the 
insulating and metallic phases are related.

The symmetry shown in Fig.~\ref{fig2} bears a strong resemblance to the 
behavior found for the resistivity near the quantum Hall liquid 
(QHL)-to-insulator transition in high mobility GaAs/AlGaAs heterostructures, 
where it has been attributed to charge-flux duality in the composite boson 
description \cite{klz}.  The symmetry was shown in this case to hold for the 
entire nonlinear $I-V$ curve \cite{shahar}.  Approximate reflection symmetry 
of the 
$I-V$ curves was also noted by van der Zant {\it et al.}\cite{vanderzant} 
at the magnetic-field-induced superconductor-insulator transition in aluminum 
Josephson junction arrays; it has been suggested that this duality can be 
traced to the symmetry between single charges in the superconducting phase and 
vortices in the insulating phase\cite{girvin}. On the other hand, there is 
no evident symmetry of the superconducting and insulating 
branches at the superconductor-insulator transition in thin films driven by 
varying thickness\cite{goldman} or a magnetic field\cite{kapitulnik}, nor do 
the $I-V$ curves show a reflection symmetry about the critical point in the 
former case\cite{goldman1}.
 
To summarize, we have presented evidence for a reflection symmetry 
about the critical point of the resistivity on one 
side and its inverse on the other side of the metal-insulator 
transition in the 2D electron gas in high mobility silicon MOSFET's in the 
absence of a magnetic field.  This implies there is a simple relation between 
the conduction mechanisms in the two phases.  The behavior near this $B=0$ 
transition is remarkably similar to that found at the quantum Hall 
liquid-insulator transition.  This suggests that some feature common to 
both transitions may be responsible for the observed duality.  A $B=0$ 
metal-insulator transition is unexpected in two dimensions, 
and its nature in high-mobility silicon MOSFET's is not currently 
understood.  The symmetry reported here may provide an additional clue that 
could lead to a theoretical understanding of the anomalous metal-insulator 
transition in 2D in the absence of a magnetic field.
	
The symmetry at the QH liquid-insulator transition provided the impetus for 
the present work: we are grateful to D. Shahar for sharing his data prior to 
publication.  We would also like to thank 
A.~M.~Goldman, S.~A.~Kivelson, D.~H.~Lee, and S.~L.~Sondhi for helpful 
discussions. This 
work was supported by the US Department of Energy under Grant No.\ 
DE-FG02-84ER45153.

\end{multicols}

\begin{references}
\bibitem{gang} E.\ Abrahams, P.\ W.\ Anderson, D.\ C.\ Licciardello, 
and T.\ V.\ Ramakrishnan, Phys.\ Rev.\ Lett.\ {\bf 42}, 673 (1979).
\bibitem{krav} S.\ V.\ Kravchenko, G.\ V.\ Kravchenko, J.\ E.\ 
Furneaux, V.\ M.\ Pudalov, and M.\ D'Iorio, Phys.\ Rev.\ B {\bf 50}, 
8039 (1994); S.\ V.\ Kravchenko, W.\ E.\ Mason, G.\ E.\ Bowker, 
J.\ E.\ Furneaux, V.\ M.\ Pudalov, and M.\ D'Iorio, Phys.\ Rev.\ B 
{\bf 51}, 7038 (1995).
\bibitem{surface} J.\ E.\ Furneaux, S.\ V.\ Kravchenko, W.\ 
Mason, V.\ M.\ Pudalov, and M.\ D'Iorio, Surf.\ Sci.\ {\bf 361/362}, 949 
(1996).
\bibitem{efield} S.\ V.\ Kravchenko, D.\ Simonian, M.\ P. \ Sarachik, W.\ 
Mason, and J. \ E. \ Furneaux, to appear in Phys.\ Rev.\ Lett. {\bf77} (1996).
\bibitem{shahar1} D.\ Shahar, D.\ C.\ Tsui, M.\ Shayegan, J.\ E.\ Cunningham, 
E.\ Shimshoni and S.\ L.\ Sondhi, to be published in Solid State Commun.
\bibitem{jain} J.\ K.\ Jain, Phys.\ Rev.\ Lett.\ {\bf63}, 199 (1989); 
Phys.\ Rev.\ B {\bf40}, 8079 (1989).
\bibitem{shahar} D.\ Shahar, D.\ C.\ Tsui, M.\ Shayegan, E.\ Shimshoni and S.\ 
L.\ Sondhi, Science {\bf 274}, 591 (1996).
\bibitem{klz} S.\ A.\ Kivelson, D.\ H.\ Lee, and S.\ C.\ Zhang, Phys.\ 
Rev.\ B\ {\bf 46}, 2223 (1992).
\bibitem{vanderzant} H.\ S.\ J.\ van\ der\ Zant, F.\ C.\ Fritschy, 
W.\ J.\ Elion, L.\ J.\ Geerligs, and J.\ E.\ Mooij, Phys.\ Rev.\ Lett.\ 
{\bf 69}, 2971 (1992), and references therein.
\bibitem{girvin} See, for example, S.\ M.\ Girvin, Science {\bf274}, 524 (1996).
\bibitem{goldman} Y.\ Liu, K.\ A.\ McGreer, B.\ Nease, D.\ B.\ Haviland, G.\ 
Martinez, J.\ W.\ Halley and A.\ M.\ Goldman, 
Phys.\ Rev.\ Lett.\ {\bf 67}, 2068 (1991).
\bibitem{kapitulnik}A.\ Yazdani and A.\ Kapitulnik, Phys.\ Rev.\ Lett.\ 
{\bf 74}, 3037 (1995).
\bibitem{goldman1} A.\ M.\ Goldman, private communication.
\end{references}
\end{document}